\documentclass[sigconf]{acmart}

\setcopyright{none}

\usepackage{xcolor,colortbl}
\renewcommand\footnotetextcopyrightpermission[1]{}

\begin{document}

\title{Survey on social reputation mechanisms: Someone told me I can trust you}

\author{Thomas Werthenbach}
\affiliation{
  \institution{Delft University of Technology}
  \city{Delft}
  \country{The Netherlands}
}
\email{T.A.K.Werthenbach@student.tudelft.nl}

\author{Johan Pouwelse}
\affiliation{
  \institution{Delft University of Technology}
  \city{Delft}
  \country{The Netherlands}
}
\email{J.A.Pouwelse@tudelft.nl}

\subtitle{--- student project ---}

\renewcommand{\shortauthors}{Thomas Werthenbach, Johan Pouwelse}

\begin{abstract}
Nowadays, most business and social interactions have moved to the internet, highlighting the relevance of creating online trust. One way to obtain a measure of trust is through reputation mechanisms, which record one's past performance and interactions to generate a reputational value. We observe that numerous existing reputation mechanisms share similarities with actual social phenomena; we call such mechanisms `social reputation mechanisms'. The aim of this paper is to discuss several social phenomena and map these to existing social reputation mechanisms in a variety of scopes. First, we focus on reputation mechanisms in the individual scope, in which everyone is responsible for their own reputation. Subjective reputational values may be communicated to different entities in the form of recommendations. Secondly, we discuss social reputation mechanisms in the acquaintances scope, where one's reputation can be tied to another through vouching or invite-only networks. Finally, we present existing social reputation mechanisms in the neighbourhood scope. In such systems, one's reputation can heavily be affected by the behaviour of others in their neighbourhood or social group.
\end{abstract}

\maketitle

\pagestyle{plain}

\section{Introduction}
Nowadays, most business and social interactions have moved to the internet, highlighting the relevance of creating online trust. The COVID-19 pandemic has shown that, in time of crises, the online news and social media usage increases \cite{van2021does}, increasing the risk and impact of misinformation. As such, it is commonly known that governments have attempted to control news media to spread propaganda in the past. Additionally, research shows that individuals getting their news from social media are often more likely to belief conspiracy theories \cite{enders2021relationship}. Such matters raise the relevant and contemporary question: who to trust?

In an automated setting, the trust measure is often extracted from one's reputation. Their reputation may be calculated through the amount of `good' work one has performed, or the reputation of their direct peers. We call systems performing such calculations \textit{reputation mechanisms}. Many reputation mechanisms have been proposed and evaluated \cite{arrep, page1999pagerank, nasrulin2022meritrank, souche, kellett2011trust, socialtrust, tian2006group, ipgrouprep}. The core components of reputation mechanisms may vary greatly, e.g. it may assume that entities have a fixed initial identity or that some entity $i$ sending some entity $j$ a message provides a proof of personhood for entity $j$. However, the common purpose of reputation mechanisms is to provide some measure of benevolence or trustworthiness.

The overall scope of this paper is focused on \textit{social reputation mechanisms}. Such reputation mechanisms are a virtual
reflection of genuine social phenomena, such as vouching or familial relationships. We provide a survey in which we have rigorously reviewed social reputation mechanisms by exploring various social concepts and mapping these to existing reputation mechanisms.

Through the course of this paper, we gradually increase our scope and consider social reputation mechanisms based on social phenomena on an increasingly larger scale. First, we discuss the individual level. In this scope, no two persons necessarily know each other initially and everyone's reputation is based solely on the work they perform or the quality they provide. Secondly, we consider one's acquaintances. Existing social ties and vouching are concepts which may transpire in this space. Lastly, we discuss phenomena occurring in one's direct neighbourhood. For instance, the neighbourhood in which you live may affect your reputation to both members outside and inside that neighbourhood.

First, we provide more background on the importance and relevance of creating trust and reputation mechanisms in section \ref{sec:background}. Section \ref{sec:definitions} provides formal definitions and data structures, which we use to generalise the mathematical foundations of reviewed mechanisms in order to reduce the usage of varying mathematical models across the different reputation mechanisms. Section \ref{sec:individual} considers entities individually, and rigorously discusses different reputation mechanisms based on social phenomena within this scope. Section \ref{sec:direct_relationships} continues exploring social concepts in the acquaintances scope and their associated social reputation mechanisms. The last scope, neighbourhoods, is discussed in section \ref{sec:neighbourhoods}. A brief overview of all discussed social reputation mechanisms can be found in Table \ref{table}.

\begin{table*}[]
\caption{
\label{table} Overview of all social reputation mechanisms reviewed in this paper, as well as associated work.}
\begin{tabular}{p{0.6cm}p{7.3cm}p{7cm}p{1.35cm}}
\textbf{Year} & \textbf{Mechanism} & \textbf{Reputation model} & \textbf{Related} \\ \hline

1999 & \textbf{PageRank} \cite{page1999pagerank}. \textit{Type: individual}. Assumes important websites are likely linked to from other websites. Links are quantified iteratively until stationary state.                       &
\begin{minipage}[t!][0.5cm][t]{0.1cm}
$$R_i = c\sum_{v \in B_i}\frac{R_i}{|N_i|}$$
\end{minipage}
            &            \cite{zhang2017pagerank, bi2008trust, wang2008research, pujol2002extracting}           \\

\rowcolor[HTML]{EFEFEF}
2006 & \textbf{GroupRep} \cite{tian2006group}. \textit{Type: neighbourhood}. Users form natural groups. The group reputation is assumed when there has not been sufficient direct interaction.      &     Given utility $u_{ij}$ and cost $c_{ij}$ from $i$ to $j$:\newline
\begin{minipage}[t!][1.1cm][t]{2cm}
$$R_{ij} = \frac{\sum c_{ij} - u_{ij}}{\sum c_{ij} + u_{ij}}$$ 
\end{minipage}&      \cite{sabater2001regret, wu2009group, sun2005adaptive, GARM, he2011social}                 \\

2008 & \textbf{BarterCast} \cite{meulpolder2008bartercast}. \textit{Type: individual}. Peers measure up- and download rates and calculate subjective reputational values using a maxflow algorithm. & Given upload rate $f_{ji}$ and download rate $f_{ij}$:\newline
\begin{minipage}[t!][1.1cm][t]{2cm}
$$R_{ij} = \frac{arctan(\gamma(f_{ji} - f_{ij}))}{\pi/2}$$
\end{minipage} & \cite{pouwelse2008tribler} \\

\rowcolor[HTML]{EFEFEF}
2009 & \textbf{IPGroupRep} \cite{ipgrouprep}. \textit{Type: neighbourhood}. Adopts IP-based groups and aggregates spam detection feedback for reputation values.                  &      Given positive and negative feedback $r_i$ and $s_i$:\newline
\begin{minipage}[t!][1.1cm][t]{2cm}
$$R_i = \frac{r_i + 1}{r_i + s_i + 2}$$
\end{minipage}
&  \cite{udmap, thomas2010rapid, ontheeffectivenessofusingip}  \\

2010  & \textbf{ARRep} \cite{arrep}. \textit{Type: individual}. Leverages direct experiences with recommendations.                               &        
\begin{minipage}[t!][0.5cm][t]{3cm}
$$R_{ij} = \alpha \cdot R_{ij}^D + (1 - \alpha) \cdot R_{ij}^R$$
\end{minipage}
&     \cite{peertrust, gauthier2004dealing, certifiedreputation, supportingtrustinvirtualcommunities, keung2008using}                  \\

\rowcolor[HTML]{EFEFEF}
2011 & \textbf{Trust by Association} \cite{kellett2011trust}. \textit{Type: acquaintances}. Invite-only network; reputation of the invitee directly affects inviter.              &         
Given some underlying reputation mechanism $U$:\newline
\begin{minipage}[t!][1.1cm][t]{2cm}
$$R_i = (1-\alpha)U_i + \alpha\frac{\sum_{j\in N_i} U_j}{|N_i|}$$
\end{minipage}  &      \cite{rogers2007disappear}                 \\

2012  & \textbf{Souche} \cite{souche}. \textit{Type: acquaintances}. Frictionless vouching and assumes all benevolent users are member of a giant connected component.                           &           Given a giant connected component (GCC), which growth is limited for each time interval: \newline \begin{minipage}[t!][1.3cm][t]{2cm}
\[
    R_i= 
\begin{cases}
    \text{Trusted},& \text{if } i \in GCC\\
    \text{Not trusted},              & \text{otherwise}
\end{cases}
\]
\end{minipage}            &                       \\

\rowcolor[HTML]{EFEFEF}

2015  & \textbf{SocialTrust} \cite{socialtrust}. \textit{Type: acquaintances}. Prefer friends over strangers. Relies on reputation of strangers if no friend available.                      &  Entity's reputation is modified based on the rating of the other party and their impact factor $T_i$:\newline
\begin{minipage}[t!][1.1cm][t]{2cm}
$$T_i = \beta\frac{R_i}{R_{max}} + (1- \beta)\frac{D_i}{D_{max}}$$ 
\end{minipage}&    \cite{kamvar2003eigentrust}                   \\

2022  & \textbf{MeritRank} \cite{nasrulin2022meritrank}. \textit{Type: individual}. Defines set of strategies to make Sybil prone reputation mechanisms Sybil tolerant.                    & Transitivity decay, connectivity decay and epoch decay applied on existing reputation mechanisms.                 &                     \\\hline  

\end{tabular}
\end{table*}
\section{Background}
\label{sec:background}
Shaping trust in the online world, arguably the telos of all reputation mechanisms, is a hard challenge, which has been studied as early as 2002 \cite{gil2002trusting, douceur2002sybil, mui2002computational}. As the space of defense mechanisms gradually evolves, so does the space of attack possibilities. For instance, people are getting more aware of the risk of the internet and start to become sceptic towards (spam)mails, causing scammers to invent more intelligent and sophisticated scams \cite{binks2019art}. Another example of the need for online trust is in the world of e-commerce, where criminals are actively attempting to swindle innocent users on large e-commerce platforms, like eBay \cite{ebayfraud}.

Nowadays, this responsibility of creating trust cannot be entrusted to private corporations. In recent events, Alphabet Inc. (e.g. Google) has been fined €220 million by French authorities for abusing its dominance in the advertisement industry. The French government has accused Google of promoting their own advertisements over their competitors'. Furthermore, in 2019, Google has been fined €1.28 billion by the European Union on similar charges \cite{googlelawsuit}. Google's dominance in the advertisement industry and the abuse of their position manifests their absolute control over the ranking of advertisements and online resources, incentivizing one to dispute their role in creating online trust. This case shows a typical example of the Red Queen hypothesis, which, in the general case, states that biological species must consistently adapt to their ever-evolving ecosystem in the eternal fight for survival \cite{redqueenbio}. In our e-commerce setting, this hypothesis corresponds to the ever-present necessity for companies to adapt to stay ahead of their evolving competition \cite{redqueen}. Such online wars only help in creating distrust between different parties, strengthening the need for widely accepted trust mechanisms. Lastly, the United States House of Representatives has composed a report assessing the power wielded by the 4 largest bigtech corporations: Facebook, Google, Amazon and Apple. These companies have become monopolies in different spaces of the internet, providing them with absolute power. It has been found that these corporations are abusing this power through ``charging exorbitant fees,
imposing oppressive contract terms, and extracting valuable data from the people and businesses that
rely on them'' \cite{449_page_rep}.

Exploiting social phenomena for the purpose of creating trust in online settings has previously been considered with the proposal of a novel peer-to-peer file-sharing system, named Tribler \cite{pouwelse2008tribler}. Tribler is a peer-to-peer file-sharing system, which introduces social ties to incentivize users not to misbehave at the expense of their friends, partners or community. Tribler suggests the usage of public and private keys as an authenticational method for recognizing previously encountered users in the anonymous peer-to-peer environment, enabling users to keep track of benevolent and malicious interactions.

\section{Terms and definitions}
\label{sec:definitions}
This section provides the formal definitions of various concepts and data structures we use in the description of existing social reputation mechanisms.

\textbf{Entity} $-$ The notion of an entity encapsulates any type of instance which may participate in the network employing the underlying social reputation mechanisms. For example, an entity may be a real person, but could also be a computer.

\textbf{Reputation mechanism} $-$ We adopt the definition of reputation mechanism as formulated by Swamynathan: ``\textit{A reputation mechanism collects, aggregates, and disseminates feedback about a user's behavior,
or reputation, based on the user's past interactions with others}'' \cite{swamynathan2010design}. In other words, a reputation mechanism processes feedback received from all entities participating in the network to cumulatively calculate a subjective or global reputation value for each entity.

\textbf{Trust} $-$ In the context of computing systems, we may adopt the definition of trust as formalized by Saputra: ``\textit{Trust is a Trustor’s level of confidence in regard to the ability of a Trustee to provide expected result in an interaction between Trustor and Trustee}'' \cite{saputra2020defining}, where a trustor is the party which receives some service and the trustee is the party entrusted with performing or providing the trustor with a certain service or resource. In other words, trust is the certainty at which entity A (trustor) believes that entity B (trustee) is able to provide them with some service. More formally, trust is defined as a weighted directional relation $(i, j, v) \in E$ between two entities $i,j \in N$ and $v \in \mathbb{R}$, where $N$ is the set of all entities, $E$ is the set of all directed relations between two entities and $v$ is the trustworthiness value assigned by  some entity $i$ to some entity $j$.

\begin{figure}
    \centering
    \includegraphics[width=0.5\linewidth]{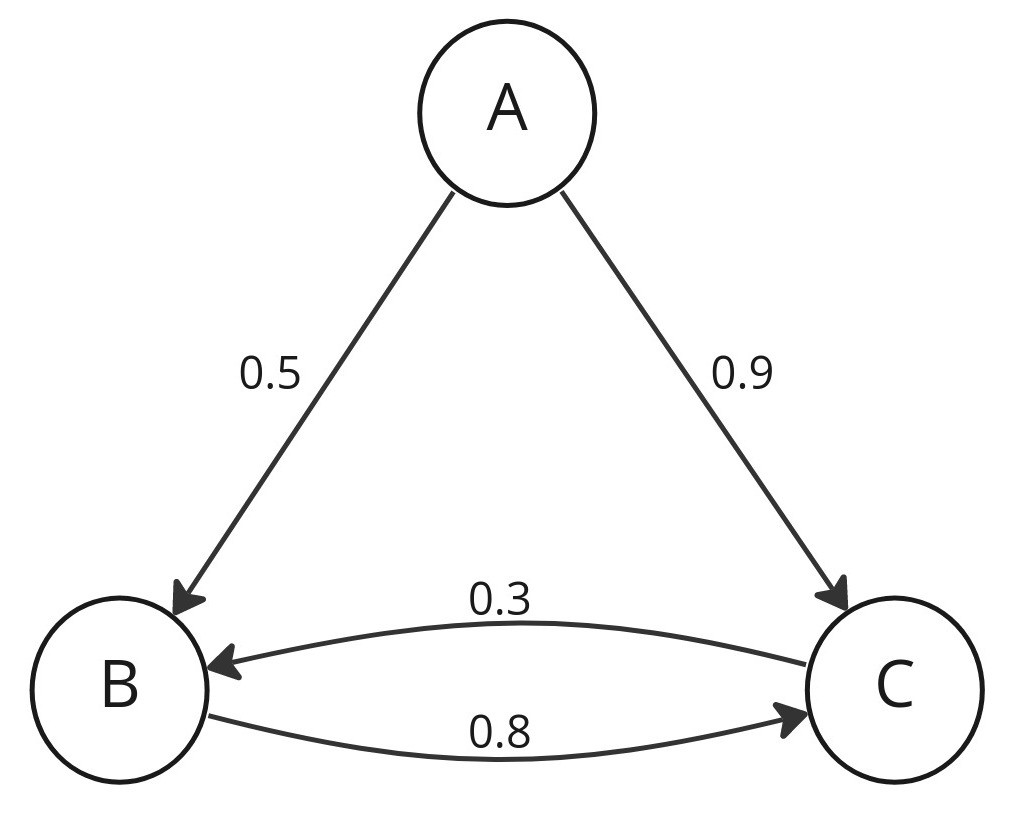}
    \caption{Example trust graph representing trust relations between entity's A, B and C.}
    \label{fig:example_trustgraph}
\end{figure}

\textbf{Trust graph} $-$ Trust relations as defined previously can be aggregated in a directed graph. We call such graph a \textit{trust graph}, or alternatively \textit{social graph}. This graph is defined by the tuple $(V, E)$, where $V$ is the set of entities and $E$ is the set of trust relationships, also referred to as edges. Such a trust graph often facilitates the necessary structural foundation. More specifically, we say that if some entity $i$ which has had sufficient (in)direct interaction with some arbitrary entity $j$, such that $j \in N_i$ and $\exists (i,j,v) \in E : v \in \mathbb{R}$, where $N_i$ is the called a \textit{trust set}, consisting of entities with whom entity $i$ has had sufficient interaction with to assess their trustworthiness, depending on the underlying reputation mechanism. Furthermore, entities can occur in multiple \textit{trust sets}, but no entity can contain itself in its trust set: $\forall i \in N : i \notin N_i$. Additionally, all entities occur at most exactly once in every trust set, such that $\forall i \in N : \{\forall j, k \in N_i : ID(j) = ID(k) \Leftrightarrow j = k\}$, where $ID$ is a deterministic implementation-specific function capable of identifying individual entities. Note that the prior implies that $\forall (i, j, v) \in E : i \neq j$. We argue that every directional relation in the graph is unique, such that $\forall (i, j, v), (k, l, w) \in E : \{(i = k \land j = l) \Leftrightarrow (i, j, v) = (k, l, w)\}$. Finally, all entities occur exactly once in a \textit{trust graph}: $\forall i,j \in N : \{ID(i) = ID(j) \Leftrightarrow i = j\}$. An example of a trust graph representing the trust relations between nodes A, B and C can be found in Figure \ref{fig:example_trustgraph}. In this example, C has a trust/reputation value of 0.8 in B's perspective, implying that $(B,C,0.8)\in E$. The reputation value is calculated and interpreted by the underlying reputation mechanism. Furthermore, B's trust set corresponds to $N_B = \{C\}$ and A's trust set to $N_A = \{B, C\}$.

\textbf{Sybil attacks} $-$ The Sybil attack \cite{douceur2002sybil} is a well-known attack used against reputation mechanisms. Many reputation mechanisms are unable to distinguish original entities from their copies \cite{levine2006survey}; a weakness abused by the Sybil attack. An adversary may employ the Sybil attack to increase its own reputation through the instant creation of virtual entities, such that they may enjoy the benefits of high reputations. The method used to increase one's reputation using `Sybil entities' depends heavily on the implementation details of the underlying social reputation mechanism. In 2011, Seuken et al. have shown that under specific circumstances, there exists a passive strongly beneficial Sybil attack \cite{seuken2011sybil}. In such an attack, a malicious entity can obtain an infinite gain with minimal effort.

\section{Individuals}
\label{sec:individual}
In the individual scope, no two entities have any initial subjective reputation value of each other and all reputations are based on the work the entities perform. However, once an entity has attained the trust of some other entity, it might propagate this trust value to peers, depending on the underlying mechanism. A physical social phenomenon resembling such situation is a networking event. During networking events, no two people have any initial measures of trust of each other, but any two people may grow to trust each other through reciprocity. The gradual creation of such trust relations may be used to form a trust graph. The assessed trustworthiness may then be shared throughout one's `network', such that someone can obtain a reputation value of an entity, which whom they did not have direct interaction. 

While direct experience with an entity is the most reliable metric to assess the trustworthiness of an entity \cite{sabater2005review}, sociological research has found that reputational values are often spread through gossip \cite{feinberg2014gossip}. Recipients of such reputational values have been shown to use these to selectively interact with cooperative rather than selfish individuals. An example of a reputation mechanism adopting this social behaviour in an online setting is ARRep \cite{arrep}.

\subsection*{ARRep}
ARRep (adaptive and robust reputation mechanism)  \cite{arrep} is a social reputation mechanism which leverages direct experience with reported experiences from other entities. While ARRep is proposed for usage in a peer-to-peer environment, the resemblance with the social phenomenon as depicted previously is vivid. Furthermore, ARRep applies  heuristic for improving the accuracy of reported experiences, by giving more weight to entities who have had more experiences.

Given some entity $i$ assessing the trustworthiness of some entity $j$, $j$'s overall reputation value $R_{ij}$ can be calculated according to:
$$R_{ij} = \alpha \cdot R_{ij}^D + (1 - \alpha) \cdot R_{ij}^R$$
where $R_{ij}^D$ represents the reputation value extracted from $i$'s direct experience with $j$, $R_{ij}^R$ corresponds to the reputation value extracted from the recommendation of peers, and $\alpha$ represents the confidence factor of $i$'s direct experience. For some threshold $M > 0$, $\alpha$ is equal to the ratio between the number of experiences and $M$ while the number of experiences is lower than $M$, otherwise $\alpha = 1$. The value of $R_{ij}^D$ corresponds to:
$$R_{ij}^D = \frac{\sum_{k = 1}^{n_{ij}} (\lambda^{n_{ij} - k} \cdot ex^k_{ij})}{\sum_{k = 1}^{n_{ij}} \lambda^{n_{ij} - k}}$$
where $n_{ij}$ is the total number of interactions between $i$ and $j$, $\lambda$ is some decay value such that $0 < \lambda \leq 1$ and $ex$ is a function returning either 1 (good) or 0 (bad) depending on the experience of interactions between $i$ and $j$ from $i$'s perspective. Moreover, the recommended reputation value $R_{ij}^R$ is calculated, such that:
$$R_{ij}^R = \frac{\sum_{i \neq k} (C_{ik} \cdot R_{ik}^D \cdot \eta^{1/n_{kj}})}{\sum_{i \neq k} C_{ik}}$$
where $\eta$ denotes some value $0 < \eta \leq 1$ and $C_{ik}$ corresponds to the recommendation credibility based on the similarity between entity $i$ and the recommender $k$ (see \cite{arrep} for details).

During evaluation, it was found that ARRep outperforms existing work \cite{peertrust} in a number of attacks for which peer-to-peer networks are susceptible. More specifically, ARRep has shown to performs better in \textit{on-off attacks}, \textit{bad mouthing attacks} and \textit{collusive cheat attacks}.\\[1\baselineskip]
There exist several reputation mechanisms similar to ARRep, focused on the same principles of combining direct experience with recommendations \cite{peertrust, gauthier2004dealing, certifiedreputation, supportingtrustinvirtualcommunities, keung2008using}. Continuing on the phenomenon in which reputation may be passed on through gossiping, an example of a reputation mechanism which directly applies this, is PageRank. PageRank uses the number of references an entity receives to determine its reputation compared to others. This behaviour is again very similar to that during networking events. PageRank has been used for assigning reputation values in social networks \cite{hogg2004enhancing} or to measure academic reputation through citation graphs \cite{massucci2019measuring}.

\subsection*{PageRank}
In the early ages of the internet, Google was among the first to adopt a reputation mechanism. Larry Page, Google's co-founder, introduced PageRank \cite{page1999pagerank}: an algorithm used to rank search engine results based on relevance. While PageRank might no longer be Google's only reputation mechanism, it is the basis of numerous other reputation mechanism \cite{zhang2017pagerank, bi2008trust, wang2008research, pujol2002extracting}.

\begin{figure}
    \centering
    \includegraphics[width=0.9\linewidth]{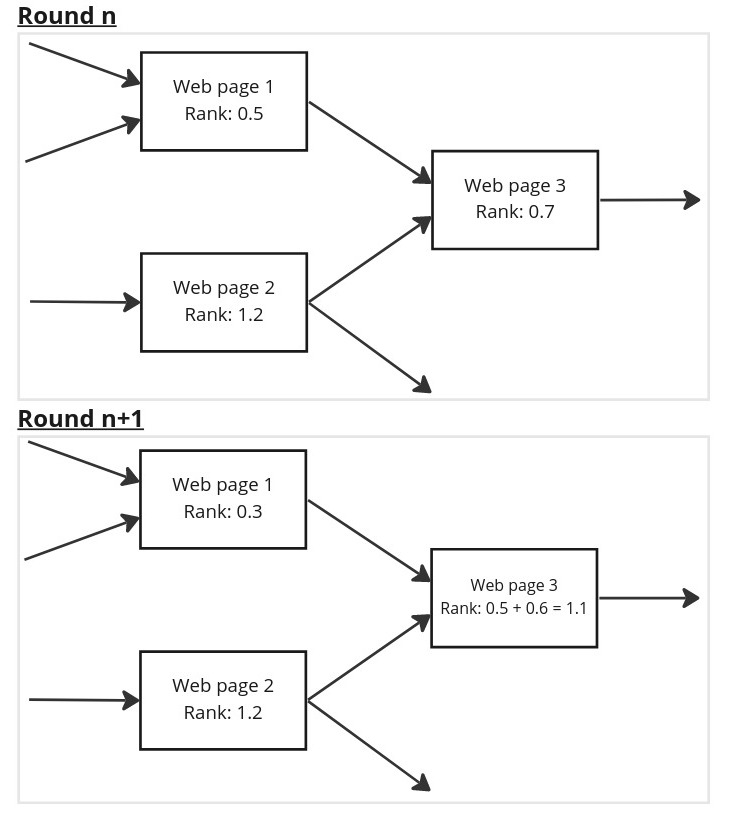}
    \caption{Two rounds of the PageRank algorithm.} 
    \label{fig:pagerank}
\end{figure}

PageRank considers the internet as a network of web pages connected through their links. If many pages link to another page, it has a higher reputation and therefore a higher `rank' on the search results page. PageRank's algorithm employs the usage of rounds: initially, every page has the same amount of `rank'. Every subsequent round, the rank flows uniformly distributed over all outgoing links to other web pages. An example execution of PageRank is displayed in Figure \ref{fig:pagerank}, in which the edges represent links between web pages. In round $n$, pages (or entities) have a specific amount of rank. In round $n+1$, the rank is propagated over the outgoing edges. Web page 3 passes the 0.7 rank over its only outgoing link while receiving 0.5 from web page 1 and $\frac{1.2}{2}$ from web page 2. Web page 1's rank is lowered and web page 2's rank is stationary. This process continues until the amount of rank for all pages becomes stationary. Once the network reaches a stationary state, extracting the final amount of rank per web page is trivial. One may note that this algorithm shows high similarity to finding the limiting probabilities of a Markov chain. 

Let A be a matrix such that $\forall (i,j,v) \in E : A_{i,j} = \frac{1}{|N_i|}$. Note that the value $v$ is not used by PageRank as it utilizes the notion of global reputation, i.e. the reputation is equivalent from all perspectives. Let R the reputation value of web page i, such that: 
$$R_i = c\sum_{v \in B_i}\frac{R_i}{|N_i|}$$
where $B_i$ is the set of states $\{j \in N\ |\ i \in N_j\}$ and c is a factor used for normalization, ensuring the total amount of `rank' remains constant. When R reaches a stationary state, i.e. it does not change anymore, it is an eigenvector of matrix A, such that $A = cAR$. However, if the trust graph takes the shape of a directed cyclic graph, loops with no outgoing edges may occur, causing the accumulation of rank over time. To tackle this issue, Page a new formula for reputational values R' of web page i, such that $R_i' = R_i + cS_i$, where $||R'||_1 = 1$ and $S_i$ is a vector of web page i which corresponds to the rank originating from each page. As we have that $||R'||_1 = 1$, c must be reduced when S is an all-positive vector, implying that c is a decay factor.

The original version of PageRank as described above is prone to Sybil attacks, as has been shown in many studies \cite{cheng2006manipulability, dinh2008sybil, danezis2006network, chang2012survey}. Such an attack would introduce many new entities who all link to the attacker, thereby increasing its reputation. This process is also known as `link farming' \cite{danezis2006network}. The original PageRank algorithm does by itself not contain any defense mechanisms against Sybil attacks. 
\\[1\baselineskip]
PageRank makes use of the notion that reputation/rank flows through a network over directed edges. However, this is not the only use case of directed edges. Another example of a use case are maxflow algorithms, in which the weights of the edges can be exploited to find the maximum \textit{flow rate} between entities. BarterCast is a reputation mechanism which makes use of such a maxflow algorithm to help determine an entity's subjective reputation and trustworthiness.

\subsection*{BarterCast}
Designed for peer-to-peer settings and deployed in Tribler \cite{pouwelse2008tribler}, BarterCast \cite{meulpolder2008bartercast} integrates up- and download rates of peers into a directed graph. By using a maxflow algorithm (e.g. Ford-Fulkerson \cite{ford_fulkerson_1957}), one can find the net up- and download rates among (in)directly connected peers. These rates are integrated through the arctan function in which -1 corresponds to a lower bound for the amount of reputation, 0 represents a neutral position (i.e. newcomers), and 1 is the upper bound on the reputation one could attain. 

More specifically, entities employing the BarterCast protocol aggregate their own up- and download speeds grouped by the entity with whom they are currently transferring files. These statistics are shared with known peers periodically, through messages known as the \textit{BarterCast messages}. Note that entities do not propagate messages they receive, implying that any given entity merely accumulates information of entities at most 2 \textit{hops} away, which is in turn used for creating a \textit{local view of the network} in the form of a directed graph. When any entity \textit{i} wants to transfer files to/from entity \text{j}, they perform a maxflow algorithm to determine the net up- and download rate between itself and entity \textit{j}, which are represented by $f_{ij}$ and $f_{ji}$, respectively. Using these values, a reputation value $R_{ij}$ is computed, such that:
$$R_{ij} = \frac{arctan(\gamma(f_{ji} - f_{ij}))}{\pi/2}$$
where $\gamma$ is a scaling factor. The usage of the arctan function has a double incentive. Firstly, it bounds the reputation such that $R_{ij} \in (-1, 1)$. Second, it ensures that changes in the net up- and download rate on a lower scale have more impact compared to a larger scale, e.g. the difference between 0KB and 100KB affects the reputation more significantly compared to 800KB and 900KB, which eases the process for newcomers. The resulting trust graph has strong similarities with real social networks, as entities can have high reputation in entity $i$'s perspective while having a low reputation in entity $j$'s point of view, depending on the network topology.

Along with a metric for reputation calculation, BarterCast is designed with a built-in resilience against the purposeful spread of false information. As all entities keep track of both their up- and downloading rate, the outcome of any correctly implemented maxflow algorithm is (upper) bounded by the statistics measured by the entity itself. However, reliance on a maxflow algorithm is accompanied with a risk for Sybil attacks, as maxflow algorithms are prone to Sybil attacks as shown by \cite{nasrulin2022meritrank}. Through parallel attacks, adversaries can trivially exploit this vulnerability and obtain infinite resources.
\\[1\baselineskip]All prior discussed reputation mechanisms are part of the family of \textit{symmetric reputation mechanisms}. In such reputation mechanisms, one's reputation only depends on the topology of the trust graph, which makes them generally prone to Sybil attacks \cite{levine2006survey}. However, such reputation mechanisms can be extended with defense mechanisms to increase their overall Sybil proofness. An example of a defensive mechanism against Sybil attacks is MeritRank, which wraps existing social symmetric reputation mechanisms and adds additional constraints, providing these mechanisms with Sybil attack tolerance.

\subsection*{MeritRank}
MeritRank \cite{nasrulin2022meritrank} is a novel reputation mechanism which main goal is to bound the gain of Sybil attacks. That is, MeritRank does not attempt to solve Sybil attacks, but merely defines a number of strategies towards tolerating them. Furthermore, MeritRank generically assumes the existence of an underlying implementation for communication and reputation calculation using a `flow-based' network, much alike the implementation used by PageRank.

Trust graphs satisfying MeritRank's constraints are shown to be Sybil tolerant. That is, for some value $0 < c < \infty$ and Sybil attack $\sigma_S$, the following holds:
$$\lim_{|S| \to \infty }\frac{\omega^+(\sigma_S)}{\omega^-(\sigma_S)} < c$$
where $S$ is the set of Sybils, $\omega^+$ is a function returning the gain for a Sybil attack and $\omega^+$ is a function returning the amount of loss for a Sybil attack. By defining certain properties for trust graph, MeritRank is capable of bounding the amount of gain an attacker can get from attacking the network. Such an attack is also known as a weakly beneficial Sybil attack \cite{stannat2021achieving}, which contrasts an attack where an adversary can obtain infinite gain, also known as a strongly beneficial Sybil attack. The constraints which MeritRank poses the trust graph are relative feedback/reputation, connectivity decay, transitivity decay and epoch decay.

The aforementioned constraints are a set of intuitive measures to bound the gain of an adversary. Relative feedback limits the amount of reputation an entity can give to some other entity by its own degree. More specifically, the updated function for assigning reputation is defined as: $$\bar{w}(i, j) = \frac{w(i,j)}{\sum_{k \in N_i} w(i, k)}$$ where $w$ is the original function for assigning reputation. Note the sum of reputation/feedback an entity assigns to its neighbours consistently equals 1. Transitivity decay defines a probability $\alpha$ which is equivalent to stop a random walk (see the Random Surfer model \cite{page1999pagerank}) for reputation determination for any given entity. Furthermore, connectivity decay defines a constant $0 \leq \beta \leq 1$ and ratio $t$, such that if for some entity $i$ (transitively) connected to some entity $j$ through some entity $k$ for at least the ratio $t$ of all possible paths, $(1-\beta)$ serves as a punishment factor for decreasing the reputation of the entity $j$ in $i$'s perspective. The connectivity decay constraint's main purpose is to identify and punish separate components. Lastly, the epoch decay defines a constant $\gamma$, which indicates the reputation decay with each epoch of the graph, incentivizing entities to keep performing work to receive reputation.

MeritRank has been evaluated on all constraints separately. It has been shown that ``transitivity decay and connectivity decay can provide a desirable level of Sybil tolerance'' \cite{nasrulin2022meritrank}. On the other hand, it was found that epoch decay, when naively implemented, may prefer new reputation assignments over existing reputation assignments. As aforementioned, MeritRank does not provide resistance against Sybil attacks, but accepts their existence and introduces a number of possible strategies towards bounding the maximum gain an attack may muster.
\\[1\baselineskip]
Individually-based social reputation mechanisms are often the most prone to Sybil attacks, as there exists no other external notion on which to base reputation calculations. While MeritRank proposes a number of strategies towards tolerating such attacks, it recognizes their existence and its inability towards preventing them. Arguably the most effective defense against Sybil attacks is the usage of fixed identity's and disabling the arbitrary creation of new virtual entities. An example of such fixed identity is the European Digital Identity \cite{eudigitalid}, which will enable EU residents to claim a single online identity. An external party verifying or providing an entity's identity is said to be the only way of preventing Sybil attacks \cite{levine2006survey}, as identities cannot instantly be created without the external party's permission and verification.

\section{Acquaintances}
\label{sec:direct_relationships}
In the scope of acquaintances, we consider social reputation mechanisms which rely on the existence of real relationships between entities. By leveraging these existing relationships, one may strengthen the defenses of online social reputation mechanisms.
An example of a social phenomenon leveraging existing relationships is \textit{vouching}: \textit{``to be able from your knowledge or experience to say that something is true''} \cite{cambridgedictionary}. In the context of reputation mechanism, vouching may generally be used as a method of putting one's reputation at stake. More specifically, in the case where some person (the \textit{voucher}) has vouched for someone else (the \textit{vouchee}), while this vouch was misplaced, the voucher loses their credibility. As a voucher willingly puts their reputation at stake for the vouchee, it makes one believe that the voucher has had prior external experience with the vouchee.

In recent years, the government of the United Kingdom has composed a rigorous guide as how to use vouches in daily-life situations \cite{govuk_2020}. It describes how people can use vouching for verifying one's identity. For instance, a parent has the ability to vouch for their child's identity. They know their child well and are certain of their child's identity, inducing no risk of vouching for them.

An example of a social mechanism employing vouching is Souche, which can be deployed on online social networks for protecting real users against fake accounts, often created for malicious purposes, such as spamming.

\subsection*{Souche}
Souche \cite{souche} is a vouch-based reputation mechanism developed partially by Microsoft\footnote{\url{https://microsoft.com/}}. Its main goal is to quickly be able to distinguish between legitimate and illegitimate users in the context of online social communities, and to slow down any malicious undetected users. Souche has been evaluated in simulations utilizing large anonymized email and Twitter\footnote{\url{https://twitter.com/}} datasets and has been shown to accurately identify 85\% of legitimate users in an early stage. Furthermore, Souche can relief users of periodic humanity checks, such as CAPTCHAs, by only performing a CAPTCHA upon registration. 

Souche's main means for creating relationships between entities, i.e. users, is through implicit vouching. Such process takes place through by considering regular activities as vouching. As such, Souche defines a vouch through emails by the conversation between two users, i.e. both users have written each other at least two emails for a conversation to be considered a vouch. Moreover, when modelling such approach to large datasets, it was found that a \textit{Giant Connected Component} (GCC) starts to take shape. Such a GCC is a large trust graph which contains 93\% of all users for the e-mail dataset, where the remaining connected components are orders of magnitude smaller than the GCC. Souche crowdsources the detection of malicious accounts, by assuming that malicious accounts are not included in the GCC.

\begin{figure}
    \centering
    \includegraphics[width=0.6\linewidth]{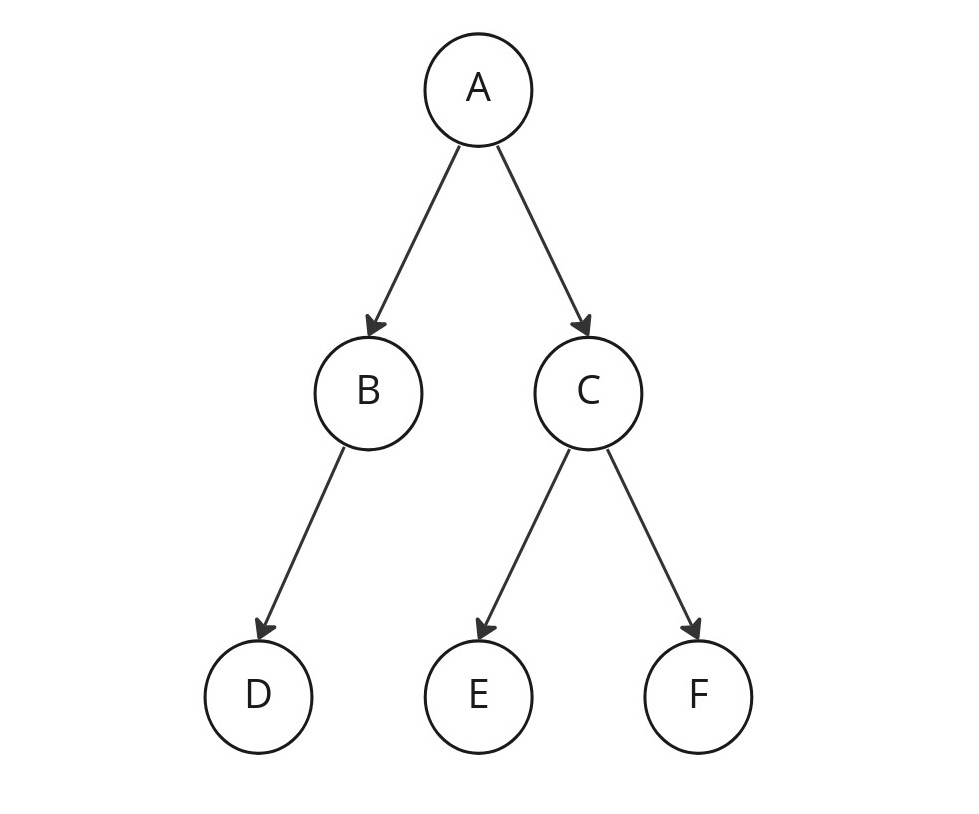}
    \caption{Example of a Souche subtree.} 
    \label{fig:souche}
\end{figure}

Souche defines a quota $q_i$ for each entity $i$ to determine whether an entity is allowed to vouch for some new entity. Every unit of time, this quota grows with rate $r$. An entity is allowed to vouch for some other entity when their quota is larger than 1. Naively, the quota can be defined as:
$$q_i = (1+r)^{t-b_i} - c_i - 1$$
where $t$ is the current time, $b_i$ is the time at which entity $i$ joined the network and $c_i$ indicates the number of entity $i$ has already vouched for. However, in order to approach the growth rate with which online social networks grow, growth rate $r$ should be configured to have a small value, such as 0.001 where the time interval equals 1 day. This implies that users are unable to vouch for any other users during their first registered year. To tackle this issue, Souche divides the GCC trust graph in subtrees, starting at the leaves, i.e. entities with no outbound vouches. An example of such a subtree can be found in Figure \ref{fig:souche}. In this particular example, A has vouched for both B and C and C has vouched for E and F. Note that if C were to be exposed as a malicious entity, it is evident that at least A, B, E and F should be further investigated, as they share a close relation to C. Souche subtrees have a size of approximately 50 entities and have a single root. Within subtrees, entities can freely use the cumulative quota. More specifically, entity $i$ of subtree $T_i$ can vouch for some other entity when $\sum_{k \in T_i} q_k > 1$. In order to account for the usage of shared quota, the definition of quota is finalized to:
$$q_i = (1+r)^{t-b_i} - c_i - d_i - 1$$
where $d_i$ represents the quota used by other entities to retain the total balance of quota within the network. Note that, due to the exponential growth of quota, older entities are assumed to be more trusted vouchers. 

Other than sharing quota, the subtree data structure serves another purpose, namely that of assisting in the detection of malicious entities. While Souche itself does not focus on malicious entity detection, given an existing detection implementation, Souche can assist by marking an entities's parent, siblings or descendants as suspicious. Another defense against malicious entities is the limited quota per time interval, preventing adversaries from vouching for other adversaries. Smaller trees will result in less available shared quota for malicious entities to claim. Finally, Sybil attacks may also suffer from these features. \\[1\baselineskip]
Another example of a study applying a vouching-based mechanism has been employed by the CloudSurfing platform \cite{couchsurfing}. This approach implements a more explicit method of vouching, requires more manual user interaction, and does not protect users from malicious and potentially fake entities, but is used as a rating for hosts on the CloudSurfing platform.

On the other hand, there exist other social phenomena leveraging existing relationships which have been translated to social reputation mechanisms. One such example includes the usage of invitations. In an offline setting, invitations are often used to invite people to participate in a certain event. This social behaviour has been studied and integrated as a core component in social reputation mechanisms, such as \textit{Trust by association}, which combines the usage of invitations with a mechanism similar to vouching.

\subsection*{Trust by association}
Trust by association (TbyA) \cite{kellett2011trust} has been designed for deployment in a peer-to-peer environment. It utilizes invitations to add new entities to the network and links the reputation values of the \textit{inviter} and \textit{invitee}, similar to vouching. More specifically, inviters may be punished for the bad behaviour of their invitees, while incentivized by profiting from the good reputation of their invitees and the rewards for growing the network. Due to these reputational incentives, it is assumed that users will only invite people they already have experience with from another channel, i.e. the acquaintance.
TbyA assumes the following properties of the network:
\begin{itemize}
    \item \textit{Invitation-only network} $-$ entities can only join the network through invitation.
    \item \textit{Homogeneous Resource or Service} $-$ entities participate in the network for a common type of resource or service.
    \item \textit{Bounded Existing Reputation Mechanism} $-$ there exists an underlying reputation mechanism, such that the resulting reputation values are bounded within a fixed interval. Kellett et al. \cite{kellett2011trust} suggests the usage of EigenTrust \cite{kamvar2003eigentrust}.
    \item \textit{Central Point of Calculation} $-$ there exists a central machine on which all calculations take place.
\end{itemize}

In its simplest form, the reputation value $R$ for entity $i$, $R_i$, is defined by:
$$R_i = (1-\alpha)U_i + \alpha\frac{\sum_{j\in N_i} U_j}{|N_i|}$$
where $U_i$ is a function returning the reputation of entity $i$ according to the underlying reputation mechanism, $N_i$ refers to the set of invitees invited by entity $i$, and $\alpha$ is some value $0 \leq \alpha < 1$ and is assumed to be 0 when $|N_i| = 0$. While \cite{kellett2011trust} only uses this formula as a starting point to introduce enhancements, the general idea remains unchanged. These enhancements include rewarding network growth by varying $\alpha$ depending on the amount of entities invited and support for recursive reputation, i.e. the reputation of the invitee's of entity $i$'s invitees affects entity.

In an effort to measure TbyA's efficacy, a simulation was performed. It was found that TbyA performs well in the case where there exists an external party capable of identifying malicious entities and punishing their inviters. TbyA is said to be able to turn lawless peer-to-peer networks into networks of benevolent peers, but requires future work on decentralized methods of identifying malicious entities. \\[1\baselineskip]
TbyA uses elements we have previously seen in Souche similar to vouching, as it punishes the inviter for any bad behaviour shown by their invitees. However, besides social reputation mechanisms in which you need a voucher to participate, there also exist less strict mechanisms. One such mechanisms is SocialTrust, in which anyone can participate, but where existing social ties are useful. SocialTrust uses the notion that friends are more trusted than strangers \cite{li2008friends}.

\subsection*{SocialTrust}
SocialTrust \cite{socialtrust} attempts to combine entity reputation values as well as friendships to provide the best QoS in a decentralized network. SocialTrust's main goal is to attempt being served by a friend or, if no friend is available, the server with the highest global reputation, provided by a \textit{trusted authority}. First of all, SocialTrust defines two types friendship, namely `friends' and `partners', both being bidirectional relationships. An entity can choose their own friends based on their experiences in the offline physical world and send a `friend request'. However, partners are assigned by the \textit{trusted authority} and are defined as entities with whom a certain entity has had many interactions with. In order to participate in a partnership, both entities must have a reputation larger than a certain partner threshold.

When some entity $i$ requires a certain service or resource, it first composes a list of all possible entities which may pose as server. In this process, the reputation mechanism takes the current load of entities into account, such that overloaded entities are not included in the list of possible servers. After composing the list, entity $i$ scans for any friends or partners and, if present, selects one of these to request the service or resource. If such friend or partner does not exist, entity $i$ queries the trusted authority for the reputations of the possible servers and chooses the server with the highest reputation.

In SocialTrust, each entity is assigned an \textit{impact factor}, which represents both their reputation the amount of damage they could inflict and is used to calculate entity's new reputation after an interaction, depending on whether it cooperates. The impact factor $T$ is defined such that:
$$T(i) = \beta\frac{R(i)}{R_{max}} + (1- \beta)\frac{D(i)}{D_{max}}$$
where $R$ is a function returning an entity's reputation, $R_{max}$ is the maximum achievable reputation, $D(i)$ represents the number of friends and partners, $D_{max}$ is the maximum number of friends and partners and $\beta$ is some value $0 \leq \beta \leq 1$. After each interaction, the client will provide a service rating of the server, which helps the trusted authority to calculate the new reputations by taking into account the impact factor.

We consider two cases: an interaction in which both the server and client are cooperative and an interaction in with the server is cooperative, but the client is non-cooperative. In the first case, the client will, subsequently to the interaction, send a service rating to the trusted authority, in which it rates the server with some value $Y$, such that $0 < Y \leq 1$. A cooperative server will accept this rating and the servers reputation increases by $\alpha(1 + T_c \cdot Y)$, where $T_c$ is the client's impact factor and $0 \leq \alpha \leq 1$. The client's reputation will increase with $\alpha$. On the other hand, we consider the cases where the client is non-cooperative and provides no feedback or negative feedback, while the server provided honest work (the trusted authority checks this by verifying the signatures on the request and response). In this case, the server is assigned $\alpha$ reputation and the client loses $-\alpha(1 + T_c)$ reputation. Similar reputation assignments are presented in \cite{socialtrust} for non-cooperative servers, in which the server loses reputation. Note that the more reputation and friends/partners an entity has, the more their reputation is affected in interactions, promoting honest work for all, regardless of reputation.

In a performance evaluation, SocialTrust has shown stronger capabilities in excluding non-cooperative entities from the network compared to EigenTrust \cite{kamvar2003eigentrust}, as well as obtaining a more accurate model mapping an entity's reputation to its benevolence. \\[1\baselineskip]
Acquaintance-based social reputation mechanism using concepts like vouching often offer built-in defenses against attacks. However, bootstrapping such mechanisms is a challenge, as they often require an initial set of trusted entities from which all remaining participants join the network. The concept of \textit{implicit vouching} as introduced by Souche might open the opportunities for deploying vouching-based mechanisms, but may inadvertently punish innocent entities. Reputation mechanisms such as SocialTrust suffer less from the bootstrap problem, but have weaker defences for filtering malicious entities.

\section{Neighbourhoods}
\label{sec:neighbourhoods}
The final scope is focused on the notion of neighbourhoods. In a social context, one's neighbourhood often determines their opportunities and success in later stages of life \cite{neighbourhoodthing}. Moreover, social groups often arise from these neighbourhoods. These groups may determine one's reputation as it has been shown that social groups are often assigned a single reputational value \cite{masuda2012ingroup}. 

Similar concepts have been applied in the design of reputation mechanisms. One such reputation mechanism is GroupRep, in which entity $j$'s reputation in entity $i$'s perspective may be determined by their group if no direct interaction has occurred.

\subsection*{GroupRep}
Based on the assumption that in large peer-to-peer networks, two peers will not often interact more than once, making it hard to profit from direct experiences between peers, GroupRep \cite{tian2006group} adopts the notion of groups to calculate reputational values. By assuming that users with similar interests in a peer-to-peer environment have constructed virtual groups, GroupRep provides a framework for calculating reputational values between groups, between groups and peers and between peers.

In GroupRep, the notion of a trust graph is applied on two scales. On the first scale, every node in the trust graph are groups of entities in which the edges represent reputations from the group perspective. The second scale considers all nodes individual entities, in which the edges represent reputation values based on direct experiences between entities. Moreover, GroupRep defines utility $u$ and costs $c$, which represent the gain and costs from interactions with other entities or entity groups. In general, reputation is calculated by $\frac{c_{ij} - u_{ij}}{c_{ij} + u_{ij}}$, where $c_{ij}$ represents the cumulative cost some entity or group $j$ has brought entity or group $i$ and $u_{ij}$ represents the cumulative utility. However, if $c_{ij} + u_{ij} = 0$, a fall-back policy is applied in which a path (on the group-based trust graph) is searched between $G(i)$ and $G(j)$, where $G$ is a function returning an entity's group. Note that for all groups along this path, including $G(i)$, the most trusted group is selected for each next step. The reputation of this path is equivalent to the minimum reputation edge on the path. However, if no such path exists, a stranger policy is applied, in which the reputation is calculated using the cumulative utility and cost for all previous interactions with strangers. Note that GroupRep will always first attempt to find direct reputation values on the trust graph on entity-level, however, if no such direct edge exists, the group reputation is used for determining a reputation value. After an interaction, entity $i$ updates its local information, creating an edge in the entity trust graph, and sends the rating to its group $G(i)$, which then may sends the rating to group $G(j)$.

Furthermore, GroupRep introduces a methodology for detecting malicious entities through clustering entities within groups. By assuming two entities as similar when they have similar reputations on the entities they both have had interactions with, clustering can take place. It is assumed that a maximum cluster of similar entities will take shape, in which all entities are deemed credible.

GroupRep has been compared against two existing reputation mechanisms on the performance against malicious collusive attacks. It was shown that GroupRep achieves a higher ratio of success queries (ratio of peers satisfied with the result of the interaction) and a higher satisfaction level, where satisfaction represents the average ratio of cumulative authentic file sizes to cumulative inauthentic file sizes. However, the scope of this evaluation was somewhat limited and did not include comparison against any well-known reputation mechanisms.\\[1\baselineskip]
While entities are still somewhat free to choose which group to join when using GroupRep, there also exist more discriminative approaches, which may be associated with originative discrimination. Such methodologies are commonly adopted in email spam measures where IP addresses are blacklisted. One such mechanism is IPGroupRep (name similarity with GroupRep is coincidental), which aggressively groups IP addresses into blocks based on subnets and assigns single reputation values to these groups based on their behaviour.

\subsection*{IPGroupRep}
IPGroupRep \cite{ipgrouprep} is an aggressive reputation mechanism for calculating a reputation for IP blocks based on existing spam classifiers. It only considers groups of IP addresses, rather than leveraging individual reputations with a group reputation. In \cite{ipgrouprep}, it is suggested to consider cluster IP into blocks of 256 by naively assuming the first 24 bits of all IP addresses in a block static, similar to a 255.255.255.0 subnet mask. An IP block's reputation should be decreased when a spam message originating from this group is detected, while it should be increased upon sending legitimate messages. Note that IPGroupRep is in itself not capable or designed to detect spam, but rather to combine the outputs of several existing spam detection mechanisms and combine these into a single reputation value.

For each group, a sum $r$ and $s$ are defined, representing the aggregation of positive and negative spam feedback respectively, provided by the numerous spam detection mechanisms. IPGroupRep applies a beta distribution, where $\alpha = r + 1$ and $\beta = s + 1$ and assumes the expected value E(p) to be the reputation value, such that:
$$E(p) = \frac{r+1}{r+s+2}$$
If this value $E(p)$ is larger than some threshold $T_{threshold}$, the group can be assumed trustworthy.

In evaluation it was found that this reputation mechanism shows very high precision and accuracy compared to existing reputation mechanisms used for the protection of mail servers. However, we argue that this method may negatively affect innocent parties within a group by disregarding the individual reputations. A possible solution to alleviate this is by decreasing the group sizes or automatically detect dynamic IP address blocks which may be used for spam \cite{udmap}.\\[1\baselineskip]
While the usage of groups may be effective against spamming and the danger of strangers, it is very generative and should be implemented cautiously such that malicious entities cannot hide in highly reputed groups and enjoy their benefits.





\section{Conclusion}
\label{sec:conclusion}
In this paper, we have discussed numerous social phenomena on different scales and reviewed social reputation mechanisms directly adopting the social phenomena as core component. First, we focused on the individual scope, in which every entity is responsible for their own reputation and entities may refer to each other based on past interactions, increasing each other's reputation by performing honest work. Secondly, we reviewed the acquaintances scope, where mechanisms may benefit from existing social ties to create more secure environments through vouching and friends. In this space, the existing trust relations are essential and may heavily influence one's reputation, compared to the individual scope. Finally, we reviewed mechanisms in the neighbourhood scope, in which entities may be part of a group which can greatly affect their reputation. Over the years, many reputation mechanisms have been proposed, evaluated and criticised. However, the holy grail of a social reputation mechanism creating secure online trust is yet to be invented.

\bibliographystyle{ACM-Reference-Format}
\bibliography{sample-base}

\end{document}